\documentclass[preprint]{ptephy_v1}

\preprintnumber{XXXX-XXXX} 
\usepackage{hyperref}
 
\usepackage{amsmath} 
\usepackage{hyperref} 
\usepackage{url} 

\usepackage{bm}
\usepackage{braket}
\usepackage{graphicx}
\usepackage{color}

\newcommand{\nn}{\nonumber}

\begin{document}

\title{An implementation of nuclear many-body wave functions by the superposition of localized Gaussians}
 

\author{Masaaki Kimura}

\affil{RIKEN Nishina Center, Wako, Saitama 351-0198, Japan}

\author[2,1,3]{Yasutaka Taniguchi}
\affil{Department of Computer Science, Fukuyama University, Fukuyama, Hiroshima 729-0292, Japan}
\affil[3]{Department of Information Engineering, National Institute of Technology (KOSEN), Kagawa College, Mitoyo, 769-1192, Kagawa, Japan}


\begin{abstract}%
  We introduce a new framework for the nuclear structure calculations, which describes the single-particle wave function as a superposition of localized Gaussians. It is a hybrid of the Hartree-Fock and antisymmetrized molecular dynamics models.
  In the numerical calculations of oxygen, calcium isotopes and $^{100}{\rm Sn}$, the framework shows its potential by significantly improving upon AMD and yielding the results consistent with or even better than Hartree-Fock(-Bogoliubov) calculations based on harmonic oscillator expansions. 
  In addition to the basic equations, general form of the matrix elements are also given.
\end{abstract}

\subjectindex{xxxx, xxx}
 
\maketitle 
  
\section{Introduction}
The nuclear many body calculations often employ density functional with local density approximation or finite-range terms. The latter offer the advantage of not only better simulating the realistic properties of nuclear forces but also the potential to use interactions derived from first principles~\cite{Feldmeier1998,Neff2003,Roth2005,Myo2015,Myo2017,Fukui2023}.
To use finite-range interactions, the wave functions are usually expanded by basis functions which are tailored to the target problems minimizing computational costs. Most commonly used basis function is the harmonic oscillator (HO) basis including deformed one~\cite{Gogny1975,Girod1983,Egido1980,Egido1995,Dobaczewski2004}. For a more accurate description of asymptotic forms of the wave functions, such as neutron halos, bases functions such as transformed HO~\cite{Stoitsov1998,Stoitsov2005} and Gaussian expansion~\cite{Nakada2006,Nakada2008} are employed. In the time-dependent approaches, the Lagrange mesh method~\cite{Hashimoto2013} has also been used. Each has its own advantages and disadvantages.

In this paper, we propose a framework which uses localized Gaussians as basis functions, which are characterized by 3-dimensional complex vectors representing their mean positions and momenta. By using them as model parameters, a variety of nuclear structure and dynamics can be described. In particular, localized Gaussian basis has advantages in describing localized cluster structures~\cite{Brink1965,Wildermuth1977,Saito1977,Horiuchi1977,Tohsaki1977} and reaction processes~\cite{Kamimura1977,Saraceno1983,Caurier1982,Drozdz1982}. 

We show that the energy variation yields two sets of equations. The first one determines the positions and momenta of localized Gaussians, which naturally extends wave packet models such as microscopic cluster models, antisymmetrized molecular dynamics (AMD)~\cite{Kanada-Enyo2003,Kanada-Enyo2012,Kimura2016}, and fermionic molecular dynamics (FMD)~\cite{Feldmeier1990,Feldmeier1995}. The second determines the coefficients of the superposition of the basis functions, which can be regarded as the Hartree-Fock equation. Thus, this framework is a hybrid of the wave packet models and Hartree-Fock.

In addition to the formulation, we demonstrate simple numerical calculations using the Gogny D1S effective interaction~\cite{Berger1991} to benchmark this framework. We calculated the ground state of spherical nuclei, i.e., oxygen, and calcium isotopes, and compared with AMD and HFB with HO basis expansion~\cite{Hilaire2007}. The present model improved AMD results, significantly in calcium isotopes. When pairing energy is negligible or not large, the present model gave equivalent or even better results compared to HF(B). The calculated proton and neutron density distributions revealed that wave functions were improved inside and outside the nucleus compared to AMD, particularly showing correct asymptotic behavior at large distances. As another example, we also present the results for $^{100}{\rm Sn}$. This nucleus is out of the applicability of ordinary AMD due to its larger mass number, but the present model also gave the reasonable results. By these examples, the present model showed its potential in studying various nuclear structure problems.

The paper is organized as follows. The next section introduces model wave function, its variational parameters and the formula for the energy variation of a simplified Hamiltonian. More general form of the Hamiltonian are discussed in the appendix. Section III demonstrates benchmark calculations for oxygen, calcium isotopes and $^{100}{\rm Sn}$. The last section summarizes this study.

\section{Framework}

\subsection{Variational wave function}
We employ a set of localized Gaussians (LGs) to describe the spatial part of the single-particle wave functions. The LG is defined as,
\begin{align}
 \braket{\bm r|p}:= \left(\frac{2\nu}{\pi}\right)^{3/4}
 \exp\Set{-\nu\left(\bm r - \frac{\bm z_p}{\sqrt{\nu}}\right)^2}, \quad
 (p=1,...,M),
 \label{eq:LG}
\end{align}
where $p$ (and $q,r,s,...$) is the index for the LGs, and $M$ is the number of the LGs which determines the size of the model space. The complex valued three-dimensional vector $\bm z_p$, which is the variational parameter, represents the average position and momentum of the LG. 
Multiplying the LG by the spin and isospin wave functions, we define the single-particle basis  functions,
\begin{align}
  \ket{p\sigma\alpha} &= \ket{p}\otimes\ket{\sigma}\otimes\ket{\alpha},
\end{align}
where the spin wave function $\ket{\sigma}$ is either $\ket{\uparrow}$ or $\ket{\downarrow}$, and the isospin wave function $\ket{\alpha}$ is either proton $\ket{p}$ or neutron $\ket{n}$.

The single-particle wave function is expressed by a superposition of these basis wave functions,
\begin{align}
  \ket{i_\alpha} &:= \sum_{p=1}^M \sum_{\sigma=\uparrow\downarrow}\ket{p\sigma\alpha}
 (c^\sigma_\alpha)_{pi}, \quad (i=1,...,Z\ {\rm or}\ N, \quad p=1,...,M), \label{eq:spwf1}\\
 c^\sigma_\alpha &:= 
 \begin{pmatrix}
  (c^\sigma_\alpha)_{11} & \cdots & (c^\sigma_\alpha)_{1Z} \\
   \vdots & \ddots & \vdots \\
   (c^\sigma_\alpha)_{M1} & \cdots & (c^\sigma_\alpha)_{MZ}
\end{pmatrix} \text{\ or\ }
\begin{pmatrix}
  (c^\sigma_\alpha)_{11} & \cdots & (c^\sigma_\alpha)_{1N} \\
   \vdots & \ddots & \vdots \\
   (c^\sigma_\alpha)_{M1} & \cdots & (c^\sigma_\alpha)_{MN}
\end{pmatrix},
\end{align}
where $\alpha$ denotes the isospin either proton or neutron, and $i$ (and $j,k,l,...$) is the index for the particle with the isospin $\alpha$, which runs upto the proton or neutron number ($Z$ or $N$). The $M\times Z$ (or $M\times N$) matrix $c^\sigma_\alpha$ is the amplitude of single-particle basis, and it is also determined by the energy variation. Thus, in this framework, the single-particle wave function has two different variational parameters, $\bm z_p$ and $c^\sigma_\alpha$.

Many-nucleon states composed of $Z$ protons and $N$ neutrons are expressed by an antisymmetrized product of the single-particle states, i.e., by Hartree-Fock approximation,
\begin{align}
  \ket{\Phi} &= \mathcal{A}\bigl [\ket{1_p}\cdots\ket{Z_p}\ket{1_n}\cdots\ket{N_n}\bigr ], \label{eq:slater}
\end{align}
where $\ket{i_p}$ and $\ket{i_n}$ denote proton and neutron single-particle wave functions.
As in the ordinary AMD for the structure calculations~\cite{Kanada-Enyo2012}, one may straightforwardly extend the model wave function by employing the parity-projected and angular-momentum-projected ones as explained in the appendix.

Let us compare the present model with others. Eq.~\eqref{eq:spwf1} can be regarded as an extension of the AMD wave function. If $Z+N=M$ and $c^\sigma_\alpha$ is diagonal, 
it reduces to the AMD wave function.  We also note that Eq.~\eqref{eq:spwf1} is similar to  the AMD-HF~\cite{Dote1997} and FMD~\cite{Feldmeier1990,Feldmeier1995} wave functions, which also describe the single-particle wave functions by the superposition of LGs.
The difference between these models and the present model lies in whether whether the LG depends on the particle index $i$ or not. For instance, Eq.~\eqref{eq:spwf1} uses the same set of LGs to describe both protons and neutrons, whereas AMD-HF and FMD do not. Hence, the present model more efficiently describes single-particle wave functions with smaller number of LGs.

In relation with the Hartree-Fock framework, Eq.~\eqref{eq:spwf1} is regarded as an implementation of Hartree-Fock by the basis expansion. However, note that the LG basis $\ket{p}$ is variational through the parameter $\bm z_p$, while ordinary Hartree-Fock formulations usually employ fixed basis wave functions. Use of variational basis function enables the present model to describe various nuclear structures with fewer basis functions as demonstrated in Sec.~\ref{sec:numerical}.

\subsection{Expectation value of the Hamiltonian}
In this work, we consider the application of this model to the nuclear structure problems. For this purpose, we evaluate the expectation value of a Hamiltonian and obtain the optimal values of the variational parameters $\bm z_p$ and $c^\sigma_\alpha$ by the energy variation.

We first introduce the overlap matrix and density matrix. The overlap matrix of LGs is defined as,
\begin{align}
 d_{pq} &:= \braket{p|q} = \exp\Set{-\frac{1}{2}(\bm z_p^* - \bm z_q)^2}, \label{eq:lgovlp}
\end{align}
and the overlap matrix of the single-particle wave functions is given by,
\begin{align}
 (b_\alpha)_{ij} := \braket{i_\alpha|j_\alpha} =
 \sum_{\sigma}\sum_{pq}d_{pq}(c^\sigma_\alpha)^*_{pi} (c^\sigma_\alpha)_{qj}.\label{eq:spovlp}
\end{align}
Note that the single-particle wave functions are unorthogonal. The density matrix is defined as in ordinary Hartree-Fock, but the inverse of the overlap matrix, $b^{-1}_\alpha$, is inserted due to the unorthogonality.
\begin{align}
 \rho^{\sigma_2\sigma_1}_\alpha := 
  c^{\sigma_2}_\alpha b_\alpha^{-1} c^{\sigma_1\dagger}_\alpha.
\end{align}
It is convenient to introduce the bi-orthonormal spinor, 
\begin{align}
 \widetilde{c}^\sigma_\alpha := c^\sigma_\alpha b_{\alpha}^{-1},
\end{align}
which simplifies the expressions of the density matrix to a form similar to the standard one,
\begin{align}
  \rho_\alpha &:= 
  \begin{pmatrix}
      \rho_\alpha^{\uparrow\uparrow} & \rho_\alpha^{\uparrow\downarrow} \\
      \rho_\alpha^{\downarrow\uparrow} & \rho_\alpha^{\downarrow\downarrow}
  \end{pmatrix}
  = 
  \begin{pmatrix}
    \widetilde{c}^{\uparrow}_\alpha \\
    \widetilde{c}^{\downarrow}_\alpha
  \end{pmatrix}
  (c^{\uparrow\dagger}_\alpha c^{\downarrow\dagger}_\alpha),
  \label{eq:dmatrix}
\end{align}
where $\rho_\alpha$ is a $2M\times 2M$ matrix. 

Now, we consider a simple Hamiltonian composed of the kinetic energy and two-body interaction. 
Let us consider the one-body kinetic energy operator $T=\sum_{i=1}^A t_i$ with $t_i = -\hbar^2\nabla^2_i/2m$. Its expectation value is given as,
\begin{align}
 \braket{T}:=\frac{\braket{\Phi|T|\Phi}}{\braket{\Phi|\Phi}} 
 = \sum_{\alpha ij}\braket{i_\alpha|t|j_\alpha}(b_\alpha^{-1})_{ji}
 = \sum_{\alpha\sigma}\sum_{pq} t_{pq}(\rho_{\alpha}^{\sigma\sigma})_{qp} \label{eq:onebody},
\end{align}
where $b_\alpha^{-1}$ appears due to the unorthogonality of the single-particle wave functions, and $t_{pq}$ is the matrix element of the kinetic energy calculated by LGs, i.e., 
$t_{pq}=\braket{p|t|q}$. 

The expectation value of a two-body potential $V = \sum_{i<j}^A v_{ij}$ is given as,
\begin{align}
 \braket{V} &:= \frac{\braket{\Phi|V|\Phi}}{\braket{\Phi|\Phi}} 
 = \frac{1}{2}\sum_{\alpha\beta}\sum_{ijkl}(b_\alpha^{-1})_{ki}(b_\beta^{-1})_{lj}
 \set{ \braket{i_\alpha j_\beta|v|k_\alpha l_\beta} 
 - \braket{i_\alpha j_\beta|v|l_\beta k_\alpha}}\nn\\
   &= \frac{1}{2}\sum_{\alpha\beta}\sum_{\sigma_1\sim\sigma_4}\sum_{pqrs}
   (\rho_\alpha^{\sigma_3\sigma_1})_{rp}(\rho_\beta^{\sigma_4\sigma_2})_{sq}
   \braket{p\sigma_1\alpha,q\sigma_2\beta|v|r\sigma_3\alpha,s\sigma_4\beta}_a,
 \label{eq:twobody1}
\end{align}
where the braket with the subscript $a$ in right hand side of Eq.~(\ref{eq:twobody1}) is the antisymmetrized matrix element of the two-body potential by LG basis functions. 
Its practial form is given in the appendix. 
By introducing the mean-field $\Gamma$, Eq.~(\ref{eq:twobody1}) is simplified as,
\begin{align}
 \braket{V} &= \frac{1}{2}\sum_{\alpha\sigma_1\sigma_3}\sum_{pr}
 (\Gamma_{\alpha}^{\sigma_1\sigma_3})_{pr}(\rho_{\alpha}^{\sigma_3\sigma_1})_{rp},
 \label{eq:twobody2}\\
 (\Gamma^{\sigma_1\sigma_3}_\alpha)_{pr} &:= \sum_{\beta \sigma_2\sigma_4}
 \sum_{qs} (\rho_\beta^{\sigma_4\sigma_2})_{sq}
 \braket{p\sigma_1\alpha,q\sigma_2\beta|v|r\sigma_3\alpha,s\sigma_4\beta}_a,
\end{align}
From Eqs.~(\ref{eq:onebody}) and (\ref{eq:twobody2}), the expectation value of the Hamiltonian is given in a form of the ordinary Hartree-Fock,
\begin{align}
  \braket{H} &:= \frac{\braket{\Psi|H|\Psi}}{\braket{\Psi|\Psi}}
    = \braket{T} + \braket{V} 
    = \sum_{\alpha\sigma_1\sigma_3}{\rm Tr} \Set{ \left(t\delta^{\sigma_1\sigma_3}
     + \frac{1}{2}\Gamma_\alpha^{\sigma_1\sigma_3}\right)\rho_\alpha^{\sigma_3\sigma_1}}.
\end{align}
In the practical calculation with Gogny density functional, one also need to incorporate with the center-of-mass correction of the kinetic energy, spin-ispspin dependent and the density-dependent two-body interactions. Different from the ordinary AMD wave function, it is impossible to separate the center-of-mass motion analytically. Hence, the center-of-mass kinetic energy is approximately removed by subtracting its expectation value.

\subsection{Energy variation}
To determine the variational parameters $c^\sigma_\alpha$ and $\bm z_p$, we consider the stationary  condition of the energy with respect to the variation of the wave function.
\begin{align}
  \delta \braket{H} &= \sum_{pi\sigma\alpha}
  \delta (c^{\sigma\dagger}_\alpha)_{ip}(\partial^{\sigma\dagger}_{\alpha})_{pi}\braket{H}
  +\sum_{p}\delta\bm z_p^*\cdot\bm \partial_p^* \braket{H} + h.c. = 0,  \label{eq:stationary}
\end{align}
where we shorthand $\partial/\partial (c^{\sigma\dagger}_\alpha)_{pi}$ and 
$\partial/\partial \bm z^*_p$ as $(\partial^{\sigma\dagger}_{\alpha})_{pi}$ and $\bm\partial_p^*$, 
respectively. From Eq.~\eqref{eq:stationary}, the stationary condition for $c^\sigma_\alpha$ is given by $(\partial^{\sigma\dagger}_{\alpha})_{pi}\braket{H} = 0$, which is calculated as,
\begin{align}
  (\partial^{\sigma\dagger}_{\alpha})_{pi}\braket{H} =
    \begin{pmatrix}
      1 - d\rho^{\uparrow\uparrow}_\alpha & -d\rho^{\uparrow\downarrow}_\alpha \\
      -d\rho^{\downarrow\uparrow}_\alpha & 1 - d\rho^{\downarrow\downarrow}_\alpha
    \end{pmatrix} 
    \begin{pmatrix}
      h^{\uparrow\uparrow}_\alpha & h^{\uparrow\downarrow}_\alpha \\
      h^{\downarrow\uparrow}_\alpha & h^{\downarrow\downarrow}_\alpha
    \end{pmatrix}
    \begin{pmatrix}
      \widetilde{c}^{\uparrow}_\alpha \\
      \widetilde{c}^{\downarrow}_\alpha
    \end{pmatrix} = 0
   \label{eq:stationary_c1}
\end{align}
with the definition of the single-particle Hamiltonian,
\begin{align}
  h^{\sigma_1\sigma_3}_\alpha := t\delta^{\sigma_1\sigma_3} + \Gamma^{ \sigma_1\sigma_3}_\alpha.
\end{align}
As explained in the appendix, Eq.~\eqref{eq:stationary_c1} is equivalent to the generalized eigenvalue equation, 
\begin{align}
  \begin{pmatrix}
    h^{\uparrow\uparrow}_\alpha & h^{\uparrow\downarrow}_\alpha \\
    h^{\downarrow\uparrow}_\alpha & h^{\downarrow\downarrow}_\alpha
  \end{pmatrix}
  \begin{pmatrix}
    \widetilde{c}^{\uparrow}_\alpha \\
    \widetilde{c}^{\downarrow}_\alpha
  \end{pmatrix} =
  \begin{pmatrix}
    d & 0 \\
    0 & d
  \end{pmatrix} 
  \begin{pmatrix}
    \widetilde{c}^{\uparrow}_\alpha \\
    \widetilde{c}^{\downarrow}_\alpha
  \end{pmatrix}
    \varepsilon, 
 \label{eq:stationary_c2}
\end{align}
where $\varepsilon = \mathrm{diag}(\varepsilon_{1},...,\varepsilon_{Z}\ \text{or}\ \varepsilon_{N})$ are the single-particle energies of protons or neutrons. Obviously, Eq.~\eqref{eq:stationary_c2} reduces to the ordinary Hartree-Fock equation, if the LGs are orthonormal, i.e., $d_{pr} = \delta_{pr}$.

The stationary condition for $\bm z_p$ is somewhat complicated but symbolically written as,
\begin{align}
  \bm \partial^*_p\braket{H} =\sum_{\alpha}\left[
  (\bm h\rho)_\alpha - \bm d(\rho h\rho)_\alpha
  \right]_{pp} = 0, \label{eq:stationary_z1}
\end{align}
where the square bracket represents an $M\times M$ matrix, and the subscript ‘pp’ denotes its $p$th diagonal element. The explicit form of these matrices are as follows:
\begin{align}
  (\bm h \rho)_\alpha &= 
  \begin{pmatrix}
    \bm h^{\uparrow\uparrow}_\alpha & \bm h^{\uparrow\downarrow}_\alpha
  \end{pmatrix}
  \begin{pmatrix}
    \rho^{\uparrow\uparrow}_\alpha\\
    \rho^{\downarrow\uparrow}_\alpha
  \end{pmatrix} +
  \begin{pmatrix}
    \bm h^{\downarrow\uparrow}_\alpha & \bm h^{\downarrow\downarrow}_\alpha
  \end{pmatrix}
  \begin{pmatrix}
    \rho^{\uparrow\downarrow}_\alpha\\
    \rho^{\downarrow\downarrow}_\alpha
  \end{pmatrix},\\
  \bm d(\rho h\rho)_\alpha &= \bm d\Set{
  \begin{pmatrix}
    \rho^{\uparrow\uparrow}_\alpha & \rho^{\uparrow\downarrow}_\alpha
  \end{pmatrix}
  \begin{pmatrix}
    h^{\uparrow\uparrow}_\alpha & h^{\uparrow\downarrow}_\alpha\\
    h^{\downarrow\uparrow}_\alpha & h^{\downarrow\downarrow}_\alpha
  \end{pmatrix}
  \begin{pmatrix}
    \rho^{\uparrow\uparrow}_\alpha\\
    \rho^{\downarrow\uparrow}_\alpha
  \end{pmatrix} +
  \begin{pmatrix}
    \rho^{\downarrow\uparrow}_\alpha & \rho^{\downarrow\downarrow}_\alpha
  \end{pmatrix}
  \begin{pmatrix}
    h^{\uparrow\uparrow}_\alpha &  h^{\uparrow\downarrow}_\alpha\\
    h^{\downarrow\uparrow}_\alpha & h^{\downarrow\downarrow}_\alpha
  \end{pmatrix}
  \begin{pmatrix}
    \rho^{\uparrow\downarrow}_\alpha\\
    \rho^{\downarrow\downarrow}_\alpha
  \end{pmatrix}},
\end{align}
where the bold-face symbols are the derivative of the single-particle Hamiltonian and the overlap matrix of LGs, which are respectively defined as,
 \begin{align}
   (\bm h_\alpha^{\sigma_1\sigma_3})_{pr} &:= 
   \bm \partial^*_p (t\delta^{\sigma_1\sigma_3})_{pr} + 
   \bm \partial^*_p (\Gamma_{\alpha}^{\sigma_1\sigma_3})_{pr}, 
   \label{eq:stationary_z2}\\
   (\bm d)_{pr} &:= \bm \partial^*_p d_{pr} = -(\bm z_p^* - \bm z_r)d_{pr}. 
   \label{eq:stationary_z3}
 \end{align}
 Note that,  in the second term of Eq.~\eqref{eq:stationary_z2}, the operator $\bm \partial^*_p$  acts only on $\bm z_p^*$ in $\braket{p\sigma_1\alpha,q\sigma_2\beta|v|r\sigma_3\alpha,s\sigma_4\beta}_a$. 
 
 Eq.~\eqref{eq:stationary_z1} determines the parameters of LGs as in the wave packet models such as AMD and FMD. Thus, the present model is a hybrid of the  Hartree-Fock and wave packet models. In the practical numerical calculations, the left hand sides of Eqs.~\eqref{eq:stationary_c1} and \eqref{eq:stationary_z1},  which give the gradient of the energy, are used in the steepest descent or conjugate gradient method to determine the optimal values of the variational parameters.

\section{Numerical results} \label{sec:numerical}
Here, we demonstrate the applicability of the present model applying to the 
 oxygen and calcium isotopes, $^{16-24}{\rm O}$ and $^{40-60}{\rm Ca}$. The Gogny D1S density functional was used for the effective Hamiltonian, and the energy variation was performed without the parity- and angular-momentum projections. The parameters $\bm z_p$ and $c^\sigma_\alpha$ were simultaneously optimized through the iterative optimization using randomly generated values as initial inputs. As the optimization algorithm, we employed the steepest descent method for $\bm z_p$ and the conjugate gradient method for $c^\sigma_\alpha$ based on the derivatives given by Eqs.~\eqref{eq:stationary_c1} and \eqref{eq:stationary_z1}.  
 The results were compared with those by the ordinary AMD with spherical Gaussian~\cite{Sugawa2001,Kimura2001} and HFB with deformed HO basis expansion~\cite{Hilaire2007}. Note that the present model does not include the pairing correlations, it underestimates the binding energies compared to HFB except for the nuclei without the pairing correlations, i.e., $^{16,24}{\rm O}$ and $^{40,48}{\rm Ca}$.

\begin{figure}[!h]
  \centering\includegraphics[width=\hsize]{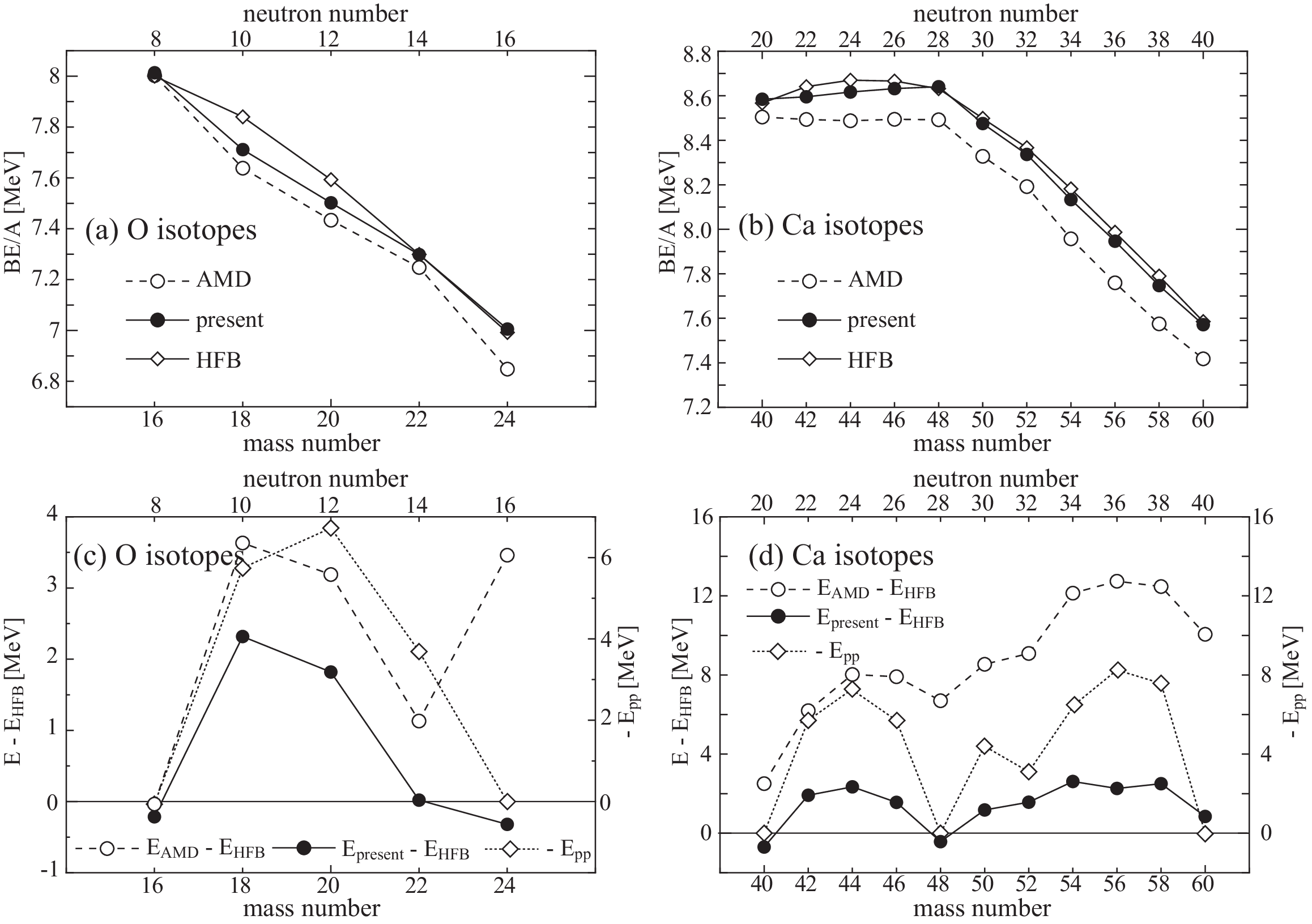}
  \caption{The calculated binding energies of oxygen and calcium isotopes in comparison with those by the ordinary AMD~\cite{Sugawa2001,Kimura2001} and HFB~\cite{Hilaire2007}. Upper panels show the calculated binding energy per particle, while the lower panels show the difference in the total energy (left vertical axis) and the particle-particle correlation energy (right vertical axis).} \label{fig:energy1} 
  \end{figure}

\begin{table}[!h]
  \caption{The optimal values of $\nu$ in the unit of $\rm fm^{-2}$ and the number of LGs, $M$, required for the convergence. The optimal values of $\nu$ were determined by a grid search with an interval of 0.0125 $\rm fm^{-2}$ in the present calculations, whereas they were determined by the steepest descent method in AMD calculations. We did not test a number of LGs less than 40 except for $^{24}{\rm O}$. }
  \label{tab:table1}
  \centering
  \begin{tabular}{cccccc}
  \hline
  nucleus & $^{16}{\rm O}$ & $^{18}{\rm O}$ & $^{20}{\rm O}$ & $^{22}{\rm O}$ & $^{24}{\rm O}$\\ 
  \hline
  $\nu$ (AMD)         & 0.158  & 0.157  & 0.156  & 0.156  & 0.149  \\
  $\nu$ (present)     & 0.2000 & 0.2000 & 0.2125 & 0.2125 & 0.1875 \\
  $M$                 & 40     & 40     & 40     & 40     & 50     \\
  \hline
  \end{tabular}
\end{table}

\begin{table}[!h]
  \caption{Same with Table.~\ref{tab:table1}, but for calcium isotopes. We did not test a number of LGs less than 60 except for $^{60}{\rm Ca}$.}
  \label{tab:table2}
  \centering
  \begin{tabular}{ccccccc}
  \hline
  nucleus & $^{40}{\rm Ca}$ & $^{42}{\rm Ca}$ & $^{44}{\rm Ca}$ & $^{46}{\rm Ca}$ 
  & $^{48}{\rm Ca}$ & \\
  \hline
  $\nu$ (AMD)     & 0.129  & 0.129  & 0.130  & 0.130  & 0.130  & \\
  $\nu$ (present) & 0.1750 & 0.1750 & 0.1750 & 0.1750 & 0.1875 & \\
  $M$             & 60     & 60     & 60     & 60     & 60     & \\
  \hline
  nucleus & $^{50}{\rm Ca}$ & $^{52}{\rm Ca}$ & $^{54}{\rm Ca}$ & $^{56}{\rm Ca}$ 
  & $^{58}{\rm Ca}$ & $^{60}{\rm Ca}$ \\
  \hline
  $\nu$ (AMD)     & 0.128  & 0.126  & 0.123  & 0.120  & 0.118  & 0.116  \\
  $\nu$ (present) & 0.1750 & 0.1750 & 0.1750 & 0.1750 & 0.1750 & 0.1750 \\
  $M$             & 60     & 60     & 60     & 80     & 80     & 80     \\
  \hline
  \end{tabular}
\end{table}

Figure~\ref{fig:energy1} shows the calculated energies of the oxygen and calcium isotopes compared with those by AMD and HFB. Except for $^{16}{\rm O}$ and $^{40}{\rm Ca}$ which are light mass and double-LS-closed shell nuclei, AMD systematically underestimated the binding energies compared to HFB. There are two reasons for this. (1) The description of the single-particle wave function is insufficient because of the restriction to the single Gaussian form. In particular, the description of $^{24}{\rm O}$ and calcium isotopes heavier than $^{48}{\rm Ca}$ is poor, which indicates the difficulty of AMD in describing the $2s_{1/2}$, $2p_{3/2}$ and $2p_{1/2}$ orbits with radial nodes. (2) The pairing correlations are not included. Therefore, difference between the AMD and HFB in the binding energies has peaks at the mid-shell nuclei in between $^{16-24}{\rm O}$,  $^{40-48}{\rm Ca}$ and $^{48-60}{\rm Ca}$ (see lower panels of Fig.~\ref{fig:energy1}).

The first problem has been resolved by the present model. The calculated energies were considerably improved, and the differences in binding energy between the present model and HFB became much smaller. The present model gave even better results (i.e., larger binding energy) than the HF(B) for $^{16,22,24}{\rm O}$ and $^{40,48}{\rm Ca}$ in which the pairing correlations are absent or not strong. Concerning the mid-shell nuclei, there still remains a difference of a few MeV, which comes from the  pairing correlations. As a reference, in the panels (c) and (d), we also plot the particle-particle correlation energy $\frac{1}{2}{\rm Tr} (\Delta\kappa^*)$ obtained by the HFB calculation in Ref.~\cite{Hilaire2007}, where $\Delta$ and $\kappa$ denote the pairing field and anomalous density, respectively. One sees that these correlation energies of mid-shell oxygen and calcium isotopes are approximately 5 to 10 MeV and reasonably explain the difference of the binding energy between the present model and HFB.
Hence, present model will yield should consistent results with the HFB calculations with the HO basis expansion, when the pairing correlations is introduced, which is now under investigation.

\begin{figure}[!h]
  \centering\includegraphics[width=\hsize]{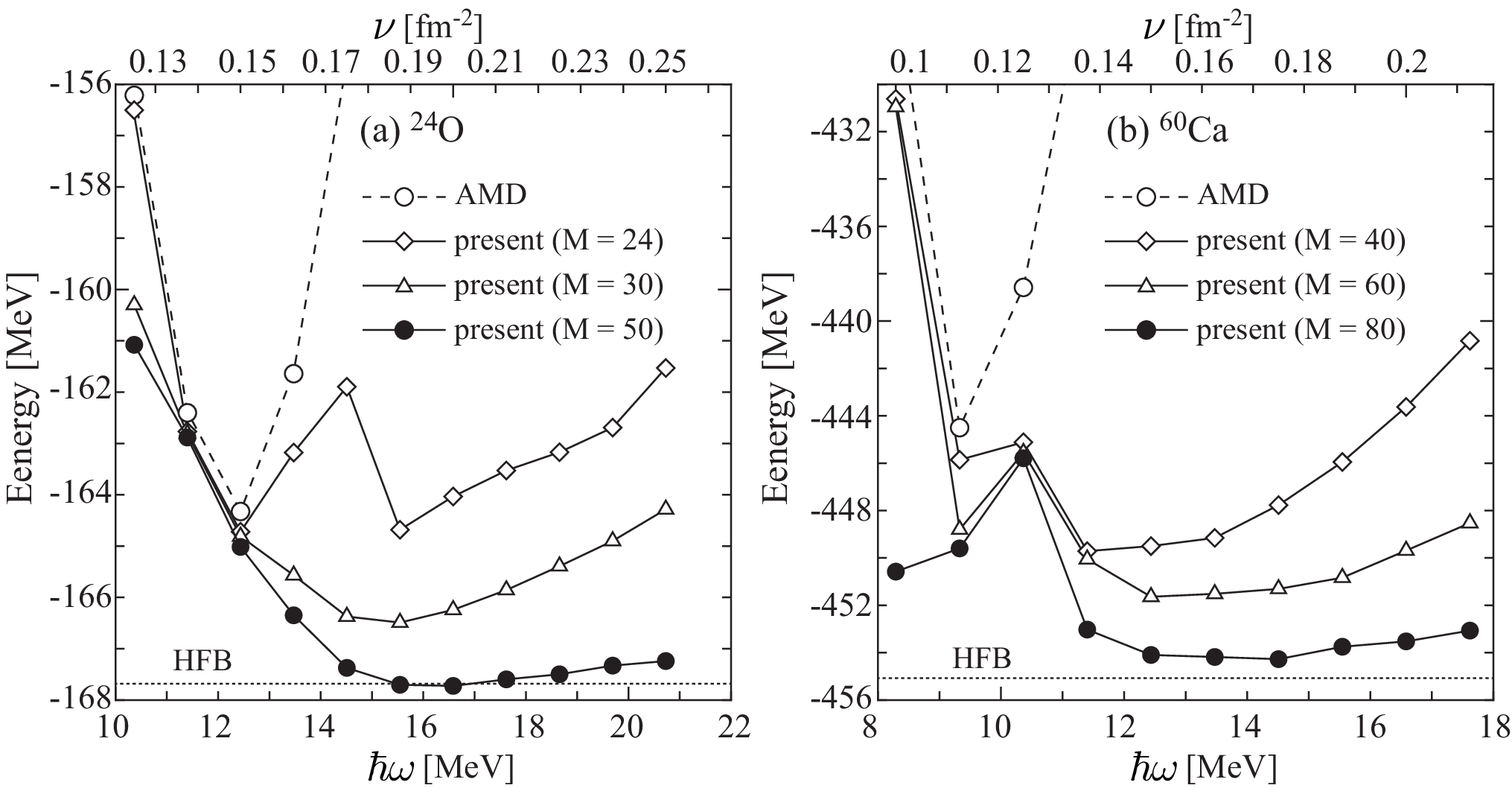}
  \caption{The calculated biding energies of $^{24}{\rm O}$ and $^{60}{\rm Ca}$ as functions of the applied Gaussian size parameter, $\hbar\omega=2\hbar^2\nu/m$. The results are plotted for the different model space with $M=24$, 30 and 50 for $^{24}{\rm O}$, and $M=40$, 60 and 80 for  $^{60}{\rm Ca}$. For comparison, the results of AMD and HFB are also shown. The results of HF(B) should also be dependent on the $\hbar\omega$ of the applied HO basis but no information was found in Ref.~\cite{Hilaire2007}.}
  \label{fig:energy2}
\end{figure}

\begin{figure}[!h]
  \centering\includegraphics[width=\hsize]{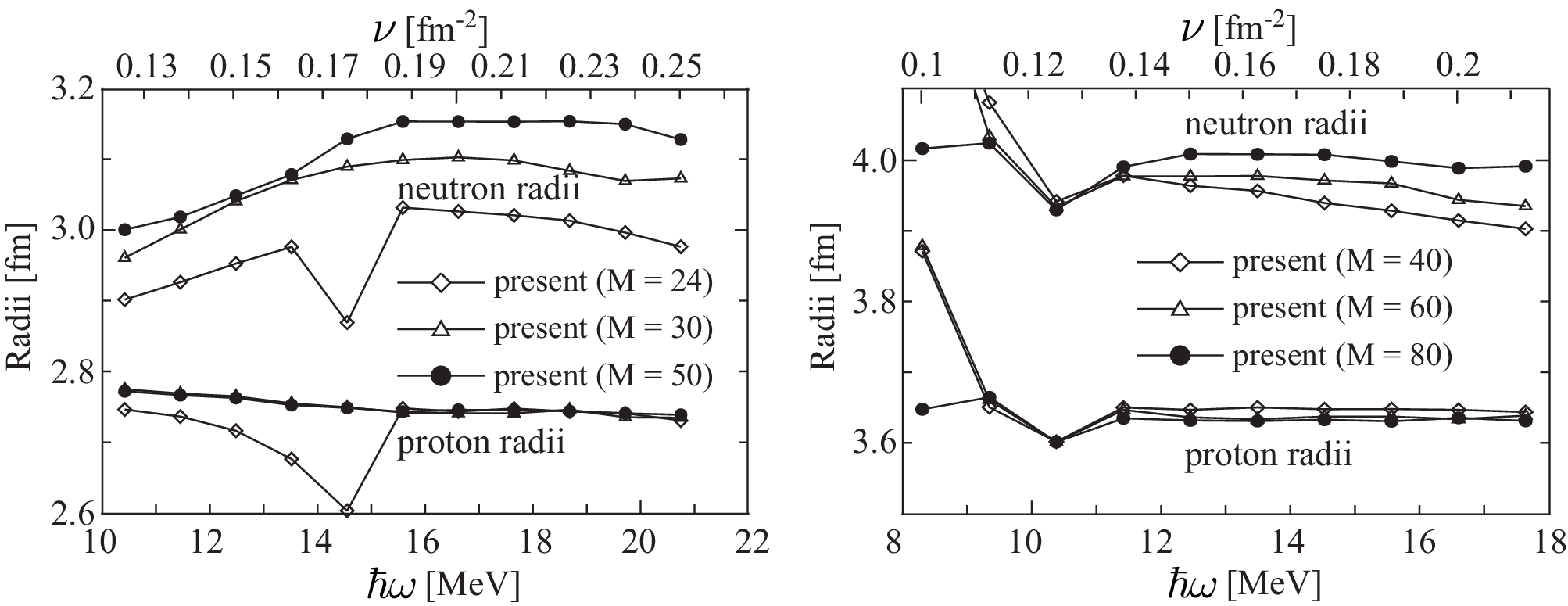}
  \caption{Same with Fig.~\ref{fig:energy2} but for the proton and neutron root-mean-square radii.}
  \label{fig:radii}
\end{figure}

\begin{figure}[!h]
  \centering\includegraphics[width=\hsize]{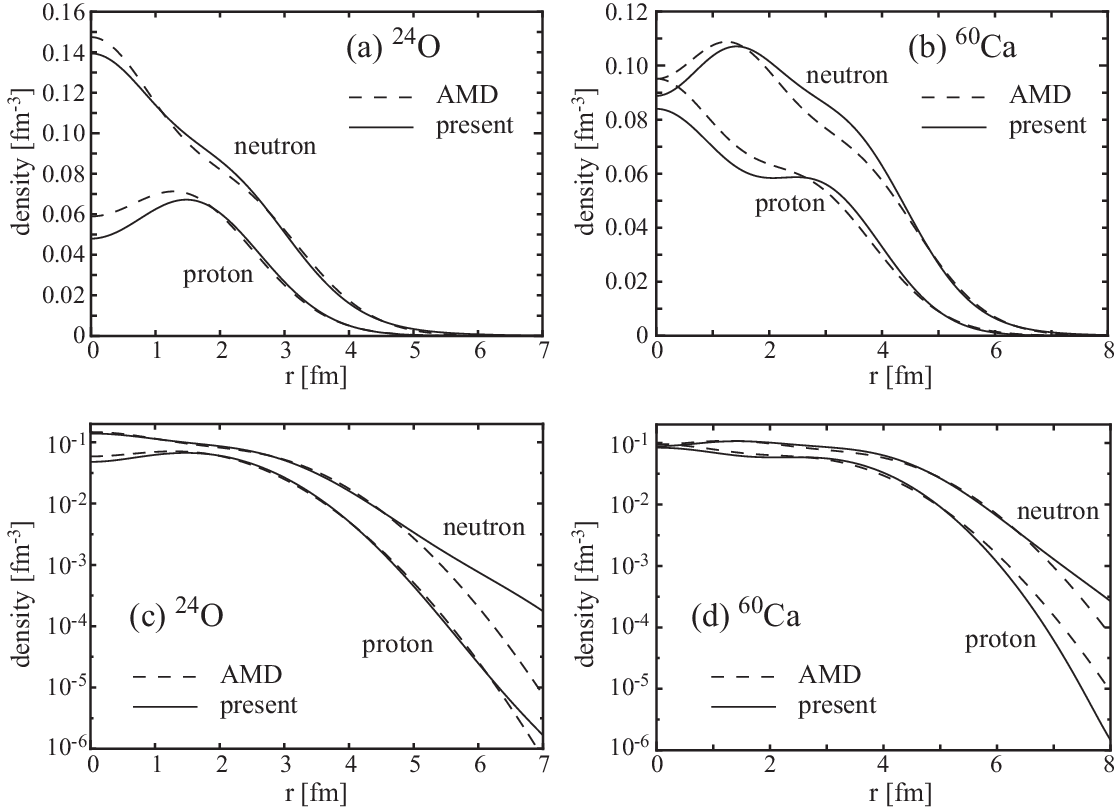}
  \caption{The calculated density distributions of protons and neutrons for $^{24}{\rm O}$ and $^{60}{\rm Ca}$ in comparison with those by the ordinary AMD. Upper and lower panels show them in the linear and logarithmic scales, respectively.}
  \label{fig_sim}
\end{figure}

Table~\ref{tab:table1} and \ref{tab:table2} show the optimized value of the width of LGs and the number of LGs required for the convergence. Compared to the AMD, the present model uses the LGs with narrower width to describe the fine structure of the single-particle wave functions. 
It is also shown that the number of LGs required for the convergence is much smaller than the 10 major shells of HO (220 basis). 
Figure~\ref{fig:energy2} shows more detailed relationship between the energy, the width and number of LGs for the cases of $^{24}{\rm O}$ and $^{60}{\rm Ca}$. 
The energy calculated by AMD is strongly dependent on the magnitude of $\nu$, because it roughly determines the size and density of nucleus. On the other hand, the present model can flexibly describe various single-particle wave functions by superposing multiple LGs. Therefore, as the number of LGs increases, the dependence of the energy on $\nu$ becomes weaker, and the energy converges smoothly. Although not shown in the figure, we have confirmed that $M>50$ for $^{24}{\rm O}$ and $M>80$ for $^{60}{\rm Ca}$ give the same converged minimum energies.
In particular, the case of $^{60}{\rm Ca}$ is impressive. The ordinary AMD underestimates the binding energy about 10 MeV compared to HFB, while the present model gave the almost the same  result with the HFB despite the lack of pairing correlations. Figure~\ref{fig:radii} illustrates the convergence for the root-mean-square radii of the obtained wave functions. As seen in the energy convergence, the dependency on the magnitude of $\nu$ decreases as the number of LGs increases sufficiently. As expected, in these neutron-rich nuclei, the proton radius converges rapidly, while the neutron radius requires larger model space for convergence.

The improvement in the wave functions is well seen from the density distributions shown in Fig.~\ref{fig_sim}. The upper panels of Fig.~\ref{fig_sim} show that the density distribution inside the nucleus has changed from that of AMD. This mainly owes to the improvement of the wave functions which have nodes inside the nucleus, such as the $2s_{1/2}$, $2p_{3/2}$ and $2p_{1/2}$. The lower panels of Fig.~\ref{fig_sim} show that the asymptotic forms outside the nucleus have also been improved. In AMD, both proton and neutron densities show Gaussian damping, and the damping rate is the same as they are described by a single Gaussian with a common value of the width parameter $\nu$. In contrast, the present model reasonably describes the exponential damping, and the damping rate is different for protons and neutrons reflecting the difference in their separation energies. In such a way, the present model greatly improves the wave functions inside and outside the nucleus.

\begin{figure}[!h]
  \centering\includegraphics[width=\hsize]{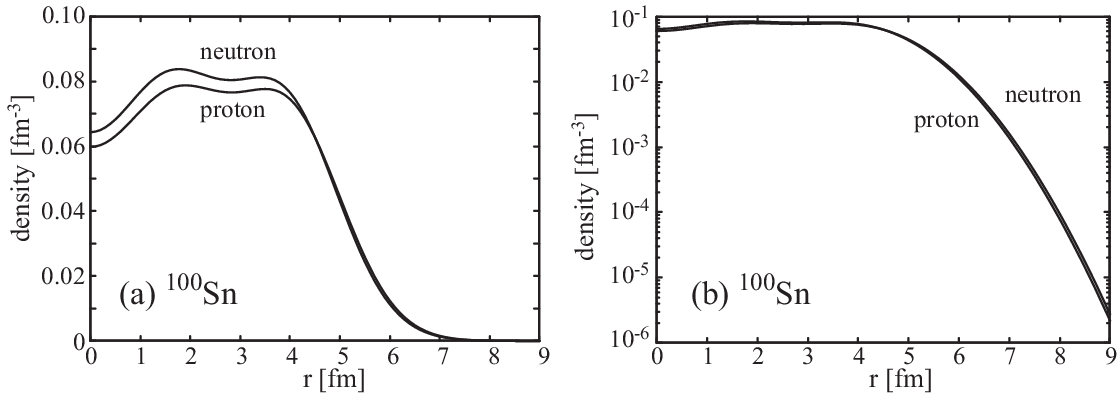}
  \caption{The calculated density distributions of protons and neutrons for $^{100}{\rm Sn}$ in linear and ogarithmic scales. The number of superposed LGs is $M=100$, and the width of LGs is set to $\nu=0.1625$ fm$^{-2}$.} \label{fig:sn}
\end{figure}

Finally, we demonstrate the applicability of the present model to the heavier mass systems. 
In the ordinary AMD, the structure calculation of such heavy nuclei was out of the scope due to the following reasons. When describing the shell structure of heavy nuclei with a limited number of LGs, the Gaussian centroids $\bm z_p$ are densely located near the center of mass increasing the overlap of LGs. Consequently, the matrix $d$ becomes ill-conditioned leading to the instability of the numerical calculations. This also makes it difficult to describe the almost constant and saturated density inside the heavy-mass nuclei. On the other hand, the present method superposes many LGs to describe the single-particle wave functions, so the overlap of LGs is not large and it is flexible enough to describe various density profiles.

As an example, for $^{100}{\rm Sn}$, the binding energy of 828.7 MeV was obtained by a superposition of a hundred LGs ($M=100$), which is 0.5 MeV deeper than the HFB with HO basis expansion~\cite{Hilaire2007}. Note that the ordinary AMD cannot give the reasonable results for this nucleus. This again confirms the efficiency of the present model. The calculated density distributions of protons and neutrons are shown in Fig.~\ref{fig:sn}. Although we cannot compare with AMD, the calculated density looks reasonable when compared with HF(B). The relatively constant density distribution inside the nucleus and the exponential damping outside the nucleus are plausibly described. The proton density distribution is slightly pushed outwards as this is a nucleus close to the proton drip line.

\section{Conclusion}

In this study, we have constructed a new framework using LGs and introduced its basic formulae. This framework describes the single-particle wave function as a superposition of LGs, where the weights of the superposition and the mean position and momentum of LGs are variational parameters. The energy variation consists of the equations determining the parameters for  LGs and the amplitudes of the superposition. The former naturally extends the motion of Gaussian wave packets described in the microscopic cluster models, AMD, and FMD, while the latter is a Hartree-Fock equation which determines the single-particle wave 
function. Since the LGs themselves are variational functions, various nuclear structure and reaction problems can be efficiently described.

In the numerical calculations of oxygen and calcium isotopes using the Gogny D1S effective interaction, the new method improved upon the results of AMD, particularly in calcium isotopes. Moreover, for nuclei in which pairing energy is not significantly large, the new method yielded the results consistent to or better than HF(B) based on HO expansions. The proton and neutron density distributions revealed the improvement of the wave functions both inside and outside the nucleus with particularly accurate asymptotic behavior at large distances. The result for $^{100}{\rm Sn}$ also demonstrated the applicability of this method to the heavier mass nuclei. Based on these results, we expect that the new method will be useful for studying various nuclear structures and reaction dynamics.

Finally, we list some extensions of the framework to be done and reported in our future works:
\begin{itemize}
  \item Use of deformed LGs~\cite{Kimura2004} to describe deformed nuclei more efficiently.
  \item Introduction of pairing correlation.
  \item Parity, angular momentum, and particle number projections; and generator coordinate method for the beyond-mean field calculations.
  \item Extenstion to time-dependent calculations for applications to low-energy nuclear reactions and real-time evolution method
  \item Introduction of nucleon collision terms for the description of high-energy nuclear reactions and nuclear fragmentation.
\end{itemize}

\section*{Acknowledgment}
The numerical computation in this work was carried out at the Yukawa Institute  (YITP) Computer Facility and at the Research Center for Nuclear Physics (RCNP) Computer System for Nuclear Physics,.
This work was supported by JSPS KAKENHI Grant Numbers JP22H01214 and JP22K03610.


\appendix

\section{Supplementary formulas}
Here, we summarize several formulae for the density matrix and the matrix element of two-body potential which are used for the energy variation. 

\subsection{The density matrix and its derivative by the variational parameters}
We remark that the density matrix defined by Eq.~\eqref{eq:dmatrix} satisfies the following identities,
\begin{align}
  N_\alpha &=\sum_{\sigma}{\rm Tr}(d\rho_\alpha^{\sigma\sigma}),\\
  \rho_\alpha &= \rho_\alpha 
  \begin{pmatrix}
    d & 0 \\
    0 & d
  \end{pmatrix}
   \rho_\alpha,
\end{align}
where $N_\alpha$ is the particle number with isospin $\alpha$. These identities reduce to the ordinary ones when LGs are orthonormal, i.e., $d_{pq}=\delta_{pq}$.

By straightforward calculations, the derivatives of the density matrix are given as,
\begin{align}
  (\partial^{\sigma\dagger}_{\alpha})_{ri}(\rho_\alpha^{\sigma_2\sigma_1})_{qp} &=
  \Set{\delta^{\sigma\sigma_1}\delta_{rp} -(d\rho^{\sigma\sigma_1}_\alpha)_{rp}}
  (\widetilde{c}^{\sigma_2}_\alpha)_{qi},\\
  \bm\partial^*_r(\rho_\alpha^{\sigma_2\sigma_1})_{qp} &=
   - \sum_\sigma(\bm{d}\rho_\alpha^{\sigma\sigma_1})_{rp}(\rho_\alpha^{\sigma_2\sigma})_{qr}.
\end{align}
Using these results, it is not difficult to derive the stationary conditions, Eq.~\eqref{eq:stationary_c1} and \eqref{eq:stationary_z1}.

\subsection{The matrix element for the two-body potentials}
Here, we give the practical expressions of the two-body mean-field potentials.
The finite-range central interaction of the Gogny force is given by the sum of two Gaussians,
\begin{align}
  v(\bm r) = \sum_{n=1}^{2}(W_n + B_nP_\sigma - H_n P_\tau - M_n P_\sigma P_\tau)e^{-r^2/\mu_n^2},
\end{align}
where $P_\sigma$ and $P_\tau$ are the spin and isospin exchange operators, respectively. For each Gaussian with the range parameter $\mu_n$, the mean-field potential is given as,
\begin{align}
  \Gamma^{\sigma_1\sigma_3}_\alpha &= \delta^{\sigma_1\sigma_3}\sum_\sigma\Bigl(
    W_n\sum_{\beta}X^{\sigma\sigma}_\beta - H_nX^{\sigma\sigma}_\alpha \Bigr)
    + B_n\sum_{\beta}X^{\sigma_1\sigma_3}_\beta - M_nX^{\sigma_1\sigma_3}_\alpha\nonumber\\
  & +\delta^{\sigma_1\sigma_3}\sum_\sigma\Bigl(
    M_n\sum_{\beta}Y^{\sigma\sigma}_\beta - B_nY^{\sigma\sigma}_\alpha \Bigr)
    + H_n\sum_{\beta}Y^{\sigma_1\sigma_3}_\beta - W_nY^{\sigma_1\sigma_3}_\alpha,
\end{align}
where $X^{\sigma_1\sigma_3}_\alpha$ and $Y^{\sigma_1\sigma_3}_\alpha$ are defined as,
\begin{align}
  (X^{\sigma_1\sigma_3}_\alpha)_{pr} &:= 
  \sum_{qs}e_{pqrs}(\rho^{\sigma_1\sigma_3}_\alpha)_{sq},\quad
  (Y^{\sigma_1\sigma_3}_\alpha)_{pr} :=
  \sum_{qs}e_{pqsr}(\rho^{\sigma_1\sigma_3}_\alpha)_{sq},\\
  e_{pqrs} &:= (1-\lambda_n)^{3/2}d_{pr}d_{qs}
  \exp\Set{-\frac{\lambda_n}{4}(\bm z_p^* - \bm z_q^* + \bm z_r - \bm z_s)^2},\\
  \lambda_n &:= (1+\nu\mu_n^2)^{-1}. \label{eq:gauss_central}
\end{align}
The total mean-field potential is obtained by summing up the contributions of all Gaussians ($n=1$,2).

The zero-range spin-orbit interaction is defined as,
\begin{align}
  v(\bm r) = 2iW_{LS} (\bm s_1 + \bm s_2)\cdot
  \overleftarrow{\nabla}\times \delta(\bm r) \overrightarrow{\nabla}.
\end{align}
For this potential, it is convenient to introduce the cartesian representation of the density matrix,
\begin{align}
  \rho^I_\alpha := \rho^{\uparrow\uparrow}_\alpha + \rho^{\downarrow\downarrow}_\alpha,\quad
  \bm \rho_\alpha := (\rho^x_\alpha, \rho^y_\alpha, \rho^z_\alpha) =
  ( \rho^{\uparrow\downarrow}_\alpha + \rho^{\downarrow\uparrow}_\alpha,\ 
   i(\rho^{\uparrow\downarrow}_\alpha - \rho^{\downarrow\uparrow}_\alpha),\ 
   \rho^{\uparrow\uparrow}_\alpha - \rho^{\downarrow\downarrow}_\alpha).
\end{align}
With this definition, the mean-field potentials in cartesian representation are given as,
\begin{align}
  \Gamma^{I}_\alpha = \sum_\beta(1+\delta_{\alpha\beta})
  \sum_{qs}\bm v_{pqrs}\cdot\bm \rho_\beta,\quad
  \bm \Gamma_{\alpha} = (\Gamma^x_{\alpha}, \Gamma^y_{\alpha}, \Gamma^z_{\alpha} ) 
  = \sum_\beta(1+\delta_{\alpha\beta}) 
  \sum_{qs}\bm v_{pqrs}\rho^I_\beta,
\end{align}
where the matrix element of the spin-orbit potential, $\bm v_{pqrs}$ is defined as,
\begin{align}
  \bm v_{pqrs} := i\nu W\left(\frac{\nu}{\pi}\right)^{3/2}
  (\bm z_p^*- \bm z_q^*)\times(\bm z_r - \bm z_s)d_{pr}d_{qs}
  \exp\Set{-\frac{1}{4}(\bm z_p^* - \bm z_q^* + \bm z_r - \bm z_s)^2}. \label{eq:zrls}
\end{align}
The spinor representation of the mean-field potentials is obtained by the following relations,
\begin{align}
  \Gamma^{\uparrow\uparrow}_\alpha &= \Gamma^I_\alpha + \Gamma^z_\alpha,\quad
  \Gamma^{\downarrow\downarrow}_\alpha = \Gamma^I_\alpha - \Gamma^z_\alpha,\quad
  \Gamma^{\uparrow\downarrow}_\alpha = \Gamma^x_\alpha - i\Gamma^y_\alpha,\quad
  \Gamma^{\downarrow\uparrow}_\alpha = \Gamma^x_\alpha + i\Gamma^y_\alpha.
\end{align}

\section{Parity and angular momentum projection}
Here, we briefly explain that the present model can be easily extended to include the parity and angular momentum projection with slight modifications. Instead of Eq.~\eqref{eq:slater}, one can use the parity-projected trial wave function,
\begin{align}
  \ket{\Phi^\pi} &= \frac{1+\pi P_x}{2}\ket{\Phi},\quad (\pi = +\ \mathrm{or}\ -)
 \end{align}
where $P_x$ denote the parity operator, and $\ket{\Phi}$ is a Slater determinant defined by Eq.~\eqref{eq:slater}. The expectation value of an operator, which is invariant under the parity transformation, is given by,
\begin{align}
  \braket{\Phi^\pi|O|\Phi^\pi} = 
  \frac{1}{2}\left\{\braket{\Phi|O|\Phi} + \pi\braket{\Phi|O P_x|\Phi}\right\}.
\end{align}
The second term of this equation can be easily calculated by applying $P_x$ to $\ket{\Phi}$, which inverts the all single-particle coordinates of the ket state as $\bm r_p\mapsto -\bm r_p$. Since the basis wave functions are the LGs, this is equivalent to chaining the sign of $\bm z_p$.
In short, $\bm z_p$ should be replaced by $-\bm z_p$ but its conjugate $\bm z_p^*$ corresponding to the bra state should not be changed. The replacement should be made in the overlap matrix of LGs (Eq.~\eqref{eq:lgovlp}) and the matrix elements dependent on $\bm z_p$ such as Eq.~\eqref{eq:gauss_central} and \eqref{eq:zrls}.

Similarly, the angular momentum-projection is also easily introduced. The trial wave function may be defined as,
\begin{align}
  \ket{\Phi^{J}_{MK}} &= \int d\Omega D^*_{MK}(\Omega)R(\Omega)\ket{\Phi},
\end{align}
where $R(\Omega)$ and $D^J_{MK}(\Omega)$ denote the rotation operator and Wigner's D function, respectively.
The expectation value of an operator, which is invariant under the rotation, is given by,
\begin{align}
  \braket{\Phi^{J}_{MK}|O|\Phi^{J}_{MK'}} = 
  \int d\Omega D^*_{KK'}(\Omega)\braket{\Phi|OR(\Omega)|\Phi}.
\end{align}
Usually, the integrals over the Euler angles are evaluated numerically. The bracket of the integrand can be directly calculated by applying $R(\Omega)$ to the ket state, which rotates both the coordinates and spin wave functions as $\bm r \mapsto R(\Omega)\bm r$ and $\ket{\sigma} \mapsto \sum_{\sigma'}D^{1/2}_{\sigma'\sigma}(\Omega)\ket{\sigma'}$. We see that they are equivalent to the back rotation of $\bm z_p$ and spinors as $\bm z_p \mapsto R^{-1}(\Omega)\bm z_p$ and $c^\sigma_\alpha \mapsto \sum_{\sigma'}D^{1/2}_{\sigma\sigma'}(\Omega)c^{\sigma'}_\alpha$, respectively. Again, the replacement should be made only for the ket state in the overlap and density matrices (Eqs.~\eqref{eq:spovlp} and \eqref{eq:dmatrix}) and in the matrix elements dependent on $\bm z_p$.

\section{Stationary condition and Hartree-Fock equation}
Here, we show that the stationary condition for $c^\alpha_{p}$ given by Eq.~\eqref{eq:stationary_c1} is equivalent to  Eq.~\eqref{eq:stationary_c2}. To this end, it is convenient to extend the single-particle wave functions to include the unoccupied (particle) states.
\begin{align}
  \ket{i_\alpha} = \sum_{p=1}^M \ket{p\sigma\alpha}(c^\sigma_\alpha)_{pi},
   \quad (i=1,...,2M, \quad p=1,...,M).
  \label{app:spwf}
 \end{align}
Note that the index $i$ runs up to $2M$\footnote{Here the factor 2 comes from the spin degrees of freedom.} over the both occupied and unoccupied states. Hereafter, for simplicity, we omit the isospin index and let $N$ as the number of particles. We assume that the index $i$ is arranged so that $i\leq N$ correspond to the occupied (hole) states and $N < i \leq 2M$ correspond to the unoccupied (particle) states. We also assume that the occupied and unoccupied states are orthogonal, i.e.,
\begin{align}
  \braket{i|j} = \sum_{\sigma}\sum_{pq}d_{pq}(c^{\sigma*})_{pi}(c^{\sigma})_{qj} = 0 \quad
  \text{for}\ i\leq N < j \leq 2M \ \text{or}\ j\leq N < i \leq 2M.
\end{align}

Accordingly, the definition of the density matrix should be renewed as
\begin{align}
  \rho &:= 
  \begin{pmatrix}
      \rho^{\uparrow\uparrow} & \rho^{\uparrow\downarrow} \\
      \rho^{\downarrow\uparrow} & \rho^{\downarrow\downarrow}
  \end{pmatrix}
  = 
  \begin{pmatrix}
    {c}^{\uparrow} \\
    {c}^{\downarrow}
  \end{pmatrix}
  \begin{pmatrix}
    b^{-1} & & &\\
    & 0& &\\
    & & b^{-1}&\\
    & & & 0
  \end{pmatrix}
  (c^{\uparrow\dagger} c^{\downarrow\dagger}).
\end{align}
where $b^{-1}$ denotes the inverse of the single-particle overlap matrix of the occupied states defined by Eq.~\eqref{eq:spovlp} ($N\times N$ matrix) whereas ‘0’ denotes the $(M-N)\times(M-N)$ zeros corresponding to the unoccupied (particle) states. Note that the stationary condition given by Eq.~\eqref{eq:stationary_c1} still holds as it is for $i\leq N$. 

Now, we confirm that $\rho d$ is the projection operator onto the occupied states. That is
\begin{align}
  &\begin{pmatrix}
    \rho^{\uparrow\uparrow} & \rho^{\uparrow\downarrow} \\
    \rho^{\downarrow\uparrow} & \rho^{\downarrow\downarrow}
\end{pmatrix}  
\begin{pmatrix}
  d & 0 \\
  0 & d 
\end{pmatrix}
\begin{pmatrix}
  c^{\uparrow}\\
  c^{\downarrow}
\end{pmatrix}
= \begin{pmatrix}
  {c}^{\uparrow} \\
  {c}^{\downarrow}
\end{pmatrix}
\begin{pmatrix}
  b^{-1} & & &\\
  & 0& &\\
  & & b^{-1}&\\
  & & & 0
\end{pmatrix}
(c^{\uparrow\dagger} c^{\downarrow\dagger})
\begin{pmatrix}
  d & 0 \\
  0 & d 
\end{pmatrix}
\begin{pmatrix}
  c^{\uparrow}\\
  c^{\downarrow}
\end{pmatrix}\nn\\
&= \begin{pmatrix}
  {c}^{\uparrow} \\
  {c}^{\downarrow}
\end{pmatrix}
\begin{pmatrix}
  b^{-1} & & &\\
  & 0& &\\
  & & b^{-1}&\\
  & & & 0
\end{pmatrix}
\begin{pmatrix}
  b & & &\\
  & \bar b& &\\
  & & b&\\
  & & & \bar b
\end{pmatrix}
= \begin{pmatrix}
  {c}^{\uparrow} \\
  {c}^{\downarrow}
\end{pmatrix}
\begin{pmatrix}
  1 & & &\\
  & 0& &\\
  & & 1&\\
  & & & 0
\end{pmatrix}
\end{align}
where $\bar b$ denotes the overlap matrix amongst the unoccupied states.
Denoting this projection operator as $P$, the stationary condition given by Eq.~\eqref{eq:stationary_c1} is rewritten as,
\begin{align}
  (1 - P)d^{-1}h\widetilde{c} = 0, \label{app:stationary_c3}
\end{align}
where we have shorthanded the $2M\times 2M$ matrices $d^{-1}$, $h$ and $\widetilde{c}$. Hence, the linear operator $d^{-1}h$ and $P$ commute. In other words, $d^{-1}h$ does not mix the occupied and unoccupied states. Because of this and the positive definiteness of the matrix $d$, there exists a  linear transformation  which rearrange $\widetilde{c}$ to satisfy the set of $2M$ generalized eigenvalue equations,
\begin{align}
  h\widetilde{c} = d\widetilde{c}\varepsilon, \label{app:stationary_c4}
\end{align}
where $\varepsilon = \mathrm{diag}(\varepsilon_1,...,\varepsilon_N,...,\varepsilon_{2M})$ are the single-particle energies of the occupied and unoccupied states.
Thus, if Eq.~\eqref{app:stationary_c3} is satisfied, then Eq.~\eqref{app:stationary_c4} is also satisfied. Conversely, if Eq.~\eqref{app:stationary_c4} is satisfied, it is obvious that Eq.~\eqref{app:stationary_c3} is also satisfied\footnote{Note that Eq.~\eqref{app:stationary_c3} holds only for the occupied states whereas Eq.~\eqref{app:stationary_c4} applies for both occupied and unoccupied states.}. Hence, the equivalence of the stationary condition and the generalized eigenvalue equations is proved.

\bibliographystyle{ptephy}
\bibliography{export}
%



\let\doi\relax








\end{document}